\documentclass[10pt,prb,aps,twocolumn,superscriptaddress,floatfix,showpacs]{revtex4}
\usepackage{graphicx}
\usepackage{amssymb}
\newcommand{\dn}{\downarrow}
\newcommand{\up}{\uparrow}
\newcommand{\half}{\hbox{$\frac{1}{2}$}}

\begin{document}

\title{Loop updates for variational and projector quantum Monte Carlo simulations \\ in the valence-bond basis}

\author{Anders W. Sandvik} 
\affiliation{Department of Physics, Boston University, 590 Commonwealth Avenue, Boston, Massachusetts 02215}

\author{Hans Gerd Evertz} 
\affiliation{Institut f\"ur Theoretische Physik, Technische Universit\"at Graz, A-8010 Graz, Austria}

\date{\today}

\begin{abstract}
We show how efficient loop updates, originally developed for Monte Carlo simulations of quantum spin systems at finite temperature, can be 
combined with a ground-state projector scheme and variational calculations in the valence bond basis. The methods are formulated 
in a combined space of spin $z$-components and valence bonds. Compared to schemes formulated purely in the valence bond basis, 
the computational effort is reduced from up to $O(N^2)$ to $O(N)$ for variational calculations, where $N$ is the system size,
and from $O(m^2)$ to $O(m)$ for projector simulations, where $m\gg N$ is the projection power. These improvements enable access 
to ground states of significantly larger lattices than previously. We demonstrate the efficiency of the approach by calculating 
the sublattice magnetization $M_s$ of the two-dimensional Heisenberg model to high precision, using systems with up to $256\times 256$ 
spins. Extrapolating the results to the thermodynamic limit gives $M_s=0.30743(1)$. We also discuss optimized variational 
amplitude-product states, which were used as trial states in the projector simulations, and compare results of projecting 
different types of trial states.
\end{abstract}


\maketitle

\section{Introduction}
\label{intro}

An ongoing challenge in simulations of quantum spin systems is to reach larger lattices sizes, thus enabling more reliable extrapolations to the 
thermodynamic limit. With the advent of loop-cluster algorithms \cite{evertz_PRL,evertz_Review,kawashima,beard,sse3} and related schemes \cite{prokofev,directed} 
developed since the mid-1990s, finite temperature ($T$) quantum Monte Carlo (QMC) simulations have become possible on lattices with millions of spins for
models with positive-definite path integral (world line) \cite{worldline,suzuki} or stochastic series expansion (SSE) \cite{sse1} representation of 
the partition function. The computational effort scales linearly in the number of spins $N$. Since the effort also scales as $1/T$, simulations at 
very low $T$ or in the ground state (using $T$ low enough to eliminate finite-$T$ effects), are limited to smaller lattices, however. The ground state 
can typically be reached for $\approx 10^4$ spins.

Currently accessible system sizes suffice for studying ground states of many important models, e.g., the two-dimensional (2D) Heisenberg model 
\cite{reger,sse2} and variants of it with non-uniform coupling patterns leading to quantum phase transitions of the antiferromagnet into a disordered 
ground state.\cite{troyer,wang} In other, similar systems there are still controversial issues \cite{wenzel,jiang1} that may need larger lattices to be 
conclusively resolved.  Larger lattice sizes are crucial in systems exhibiting more complex ground states and quantum phase transitions. One example 
of current interest is a class of ``J-Q'' models---2D Heisenberg models with four-spin interactions engineered to destroy the antiferromagnetic order 
and drive the system into a valence-bond-solid (VBS) state.\cite{deconf,melko,jiang2} The VBS state has intricate fluctuations and the true nature of its 
thermodynamic limit is only manifested on large lattices.\cite{deconf} This illustrates the need to develop better ground-state methods, as a more efficient 
alternative to going to very low $T$ with finite-temperature methods.

In this paper we introduce a method combining loop updates first developed for finite-$T$ simulations \cite{evertz_PRL,evertz_Review,sse3} with a 
ground-state projector QMC method operating in the valence bond (VB) basis.\cite{vbmethod1,vbmethod2} This over-complete singlet basis has some features 
that make it uniquely well suited for studies of spin-rotationally invariant hamiltonians such as the Heisenberg model and its extensions with multi-spin 
interactions.\cite{deconf}

It has been known for some time \cite{evertz_Review} that there is a simple and 
elegant relationship between VB states consisting of $N/2$ pairs of spins forming singlets,\cite{nachtergale} and the loop algorithm, which indeed works by 
switching between a VB basis and a basis of $N$ spins $\up$ and $\dn$ (for $S=\half$ systems). Here we exploit this switching for 
ground state projections. An attractive feature of this approach is that it enables the use of very good singlet trial wave functions, the simplest example 
of which is the amplitude-product state proposed by Liang, Doucot and Anderson.\cite{liang1,jievar} The ground state can then be reached much faster than 
with finite-$T$ methods, and with much less computational effort than projector methods formulated purely in the VB basis.\cite{liang2,santoro,vbmethod1}
In addition we show that purely variational calculations can also be made more efficient by combining spins and VBs, including a loop update similar to one previously developed for
classical dimer models.\cite{adams,sandvikdimer} 

The projector QMC algorithm with loops, working in the combined space of VBs and spins, is in the end very similar to $T>0$ SSE and worldline loop
algorithms. Essentially, the $T=0$ projector approach corresponds to ``cutting open'' the periodic imaginary-time boundary and ``sealing'' the resulting
open loop segments with valence bonds (which serve as continuations of the loops). 

We demonstrate the efficiency of the projector method by producing 
high-precision bench-mark results for the sublattice magnetization of the 2D Heisenberg model with up to $256\times 256$ spins. We also discuss the
properties of the variationally optimized amplitude-product states used as a trial states for the ground-state projections (extending the results
of Ref.~\onlinecite{jievar} to larger lattices).

We begin in Sec.~\ref{basis} by summarizing the properties of the VB basis needed for formulating the algorithms. In Sec.~\ref{variational} we
discuss variational Monte Carlo simulations and optimization of amplitude-product states, and in Sec.~\ref{projector} we describe the projector QMC 
method. We show results of both variational and projector calculations in Sec.~\ref{results}, and conclude in Sec.~\ref{summary} with a summary
and discussion.

\section{The valence bond basis}
\label{basis}

A VB state is a product of two-spin singlets,
\begin{equation}
(a,b)=(|\up_a\dn_b\rangle-|\dn_a\up_b\rangle)/\sqrt{2},
\label{singlet}
\end{equation}
where $a$ and $b$ denote sites on sublattice $A$ and $B$ on a bipartite system. For $N$ (even) spins there are $(N/2)!$ ways to draw the VBs, and hence 
the basis is massively overcomplete. The properties of VB states have been discussed extensively in the literature.\cite{liang1,sutherland,beach}
Here we will only summarize the most important aspects of the basis and introduce some notation needed for our algorithms. 

\begin{figure}
\includegraphics[width=6cm, clip]{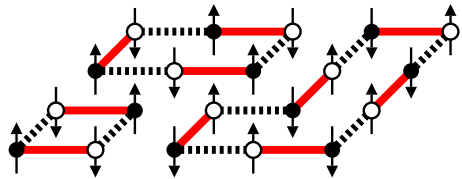}
\vskip-2mm
\caption{(Color online) Transposition graph of VB states on a $5\times 4$ lattice with sublattices A and B indicated with open and solid circles.
Bond configurations of a bra $\langle V_l|$ and ket state $|V_r\rangle$ are shown as dashed and solid lines, respectively. For clarity, all bonds here 
are of length one lattice spacing, but in general a bond can connect any pair of sites on different sublattices. The number of loops is $N_{\rm loop}=3$ 
and the number of sites $N=20$, giving, according to Eq.~(\ref{overlap1}), an overlap  $\langle V_l|V_r\rangle=2^{-7}$. A configuration of $\up$ and $\dn$ 
spins compatible with both VB states is also shown.}
\label{fig1}
\vskip-3mm
\end{figure}

We can formally use a label $r=1,\ldots,(N/2)!$ for enumerating the VB configurations and denote a state as
\begin{equation}
|V_r\rangle = |(a_1^r,b_1^r)(a_2^r,b_2^r)\cdots(a_{N/2}^r,b_{N/2}^r)\rangle.
\end{equation}
The overlap between two VB states is
\begin{equation}
\langle V_l |V_r\rangle = 2^{N_{\rm loop}-N/2},
\label{overlap1}
\end{equation}
where $N_{\rm loop}$ is the number of loops formed in the so-called {\it transposition graph} when the bonds in $|V_l\rangle$ and $|V_r\rangle$ are 
superimposed. An example is shown in Fig.~\ref{fig1}. 

Like the overlap, matrix elements of operators of interest can typically also be related to the transposition-graph 
loops.\cite{sutherland,liang1,beach} To compute spin-spin correlations we will need
\begin{equation}
\frac{\langle V_l|{\bf S}_i\cdot {\bf S}_j|V_r\rangle}{\langle V_l|V_r\rangle}=
\left\lbrace \begin{array}{ll}
0, &    {\rm if~}\lambda_i\not = \lambda_j, \\
\phi_i\phi_j(3/4), & {\rm if~}\lambda_i= \lambda_j, 
\end{array}\right.
\label{ssestimator}
\end{equation}
where $\phi_i=\pm 1$ for sites on sublattice A and B, respectively, and $\lambda_i \in \{1,N_{\rm loop}\}$ is a label distinguishing the loop 
to which site $i$ belongs. More complicated matrix elements and their relationships to the loop structure are discussed in Ref.~\onlinecite{beach}.

Fig.~\ref{fig1} also shows one of the spin states,
\begin{equation}
|Z^r_i\rangle = |S^z_{1}(r,i),\ldots,S^z_{N}(r,i)\rangle,~~~ i=1,\ldots,2^{N/2},
\label{zri}
\end{equation}
contributing to both the VB states (meaning that the spins on all bonds of the transposition graph are anti-parallel). In (\ref{zri}) the index $i$ 
refers to the allowed spin states in $|V_r\rangle$; the $2^{N/2}$ states with anti-parallel spins on every bond. With the sign convention in 
(\ref{singlet}), the VB state can be written as
\begin{equation}
|V_r\rangle = \frac{1}{2^{N/4}}\sum_{i=1}^{2^{N/2}}(-1)^{A_{\up}(r,i)}|Z^r_i\rangle, 
\label{vrz}
\end{equation}
where $A_{\up}(r,i)$ denotes the number of $\up$ spins on sublattice $A$. The VB states thus obey the well known Marshall sign rule, which
ensures that the ground state has a positive-definite expansion in terms of them [for an interaction with the same A-B bipartite structure as
in the definition of the valence bonds, Eq.~(\ref{singlet})]. 

When writing the overlap (\ref{overlap1}) using spin states,
\begin{equation}
\langle V_l |V_r\rangle = \frac{1}{2^{N/2}}\sum_{i,j}\langle Z^l_j|Z^r_i\rangle (-1)^{A_{\up}(r,i)+A_{\up}(l,j)},
\label{overlap2}
\end{equation}
only the terms with $Z^r_i=Z^l_j$ contribute. Since the spins on every bond must be anti-parallel, the spin configuration around a loop in the overlap 
graph must be staggered. The signs cancel and the sum then simply counts the number of such configurations. There are two staggered spin patterns for 
each loop, resulting in the overlap (\ref{overlap1}). Here we will use the equivalence between the two ways of expressing the overlap, (\ref{overlap1}) 
and (\ref{overlap2}), as a starting point for constructing efficient variational and ground-state projector algorithms.

\section{Variational Monte Carlo}
\label{variational}

We first discuss variational calculations, which we will use to construct good trial states for subsequent ground-state projection.
An arbitrary singlet state $|\Psi\rangle$ can be expanded in VB states;
\begin{equation}
|\Psi\rangle = \sum_{r}w_r |V_r\rangle.
\label{psivb}
\end{equation}
Because of the over-completeness, the expansion coefficients $w_r$ are not unique, but this is not a problem in practice. In the variational 
state introduced by Liang {\it et al.},\cite{liang1} the coefficients are taken to be products of bond amplitudes 
$h({\bf r})>0$,
\begin{equation}
w_r=\prod_{{\bf r}}h({\bf r})^{N_r({\bf r})},
\label{wamp}
\end{equation}
where $N_r({\bf r})$ is the number of bonds of size ${\bf r}$, where in a translationally-invariant system, all applicable lattice symmetries can 
be used, e.g., for a 2D square lattice the amplitudes are $h(x,y)$, with $x=|r_x|$ and $y=|r_y|$ (the components of the vector
${\bf r}$ defining the ``shape'' of the bond, transformed to the all-positive quadrant), and also $h(y,x)=h(x,y)$.

To optimize the amplitudes using variational QMC simulations, the energy is written as
\begin{equation}
E = \frac{\langle \Psi | H |\Psi\rangle}{\langle \Psi |\Psi\rangle} =
\frac{\sum_{lr}W_{lr}E_{lr}}{\sum_{lr}W_{lr}},
\end{equation}
where the weight $W_{lr}$ and energy estimator $E_{lr}$ are
\begin{equation}
W_{lr}=w_l w_r \langle V_l |V_r\rangle,~~~~E_{lr}=\frac{\langle V_l | H |V_r\rangle}{\langle V_l |V_r\rangle}.
\end{equation}
For a model with two-spin interactions (to which the methods discussed here are not limited,\cite{deconf,vbmethod1} but for simplicity of the discussion we will 
not consider higher-order interactions here), the energy estimator is a sum of terms of the form (\ref{ssestimator}). As in any variational calculation, the 
idea is to compute $E$ and its derivatives with respect to the amplitudes and then use this information to periodically adjust the amplitudes, in order to 
approach the  energy minimum. Here we will not be concerned with the optimization aspect of the problem (which is discussed in detail in Ref.~\onlinecite{jievar}, 
including also explicit expressions for evaluating the derivatives), but focus on the Monte Carlo sampling aspects of the problem.

\subsection{Monte Carlo sampling}

\begin{figure}
\center{\includegraphics[width=8.4cm, clip]{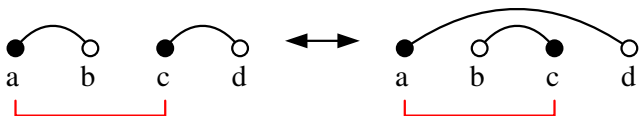}}
\vskip-2mm
\caption{(Color online) A two-bond update. For any pair of two sites on the same sublattice (here the sites a, c, as indicated by the brackets), 
the two bonds connected to them can be reconfigured in a unique way while maintaining the restriction of bipartite bonds. Detailed balance is then
satisfied for updates toggling between the two configurations, with the Metropolis acceptance probability (\ref{paccmetro}).}
\label{fig2}
\vskip-3mm
\end{figure}

For a given set of amplitudes, the VB configurations can be sampled according to the weight $W_{lr}$ by simple moves of bond pairs.\cite{liang1} Choosing 
two sites on the same sublattice (e.g., next-nearest-neighbors on the square lattice), there are two allowed (bipartite) configurations of the two bonds 
connected to these sites, and the Monte Carlo move amounts to switching from the current one to the other possible one. Such an update is illustrated
in Fig.~\ref{fig2}. The ratio $W_{l'r'}/W_{lr}$ of the weights after ($W_{l'r'}$) and before ($W_{lr}$) the update  needed for the Metropolis 
acceptance probability;
\begin{equation}
P_{\rm accept} = {\rm min}\left [\frac{W_{l'r'}}{W_{lr}},1 \right ].
\label{paccmetro}
\end{equation}
The ratio involves the amplitudes of the affected bonds from the amplitude product (\ref{wamp}), as well as the ratio of the new to old overlap 
(\ref{overlap1}). Here the number of loops $N_{\rm loop}$ can increase or decrease by one. Determining this change requires tracing a loop going through 
one of the four sites involved in the two-bond move. Starting from one of these sites, the loop going through that site before the re-configuration 
is followed, until some other site out of those four is reached. If that site is connected to the same bond as the initial site, then the four sites must 
belong to two different loops (and after the move they will all be in the same loop, i.e., $N_{\rm loop}$ is decreased by $1$), whereas if it belongs 
to the second bond all four sites are in a single loop (which is split into two after the update, leading to $N_{\rm loop}$ increasing by $1$). The loop tracing
is the most time consuming part of the calculation, in particular for systems where the loops are typically long. The loop lengths are related to the squared 
sublattice magnetization, according to\cite{sutherland}
\begin{equation}
M_s^2 = \frac{1}{N^2} \sum_{ij}\phi_i\phi_j {\bf S}_i\cdot {\bf S}_j = \frac{1}{N^2}\sum_{\lambda=1}^{N_{\rm loop}}L_\lambda^2,
\end{equation}
where $L_\lambda$ is the length of loop $\lambda$. 
For the single loop of length $L_i$ going through a randomly chosen site $i$, this relation becomes \cite{wolff,evertz_Review}
\begin{equation}
 \left\langle L_i \right\rangle = N \left\langle M_s^2 \right\rangle ,
\end{equation}
where the averages refer to the Monte Carlo sampling. For systems with antiferromagnetic long-range order there are system-spanning 
loops,\cite{sutherland,liang1} i.e., there are some loops of length $L_\lambda \propto N$ in typical configurations. In this case, the computational
effort of a sweep of $\propto N$ two-bond updates therefore scales as $N^2$.

Here we avoid the loop-tracing step by re-introducing the spins into the sampling space, replacing the pure-bond overlap (\ref{overlap1}) with the equivalent 
expression (\ref{overlap2}). In addition to sampling the bond tilings $V_l,V_r$ of the bra and ket states, we then also have to sample their compatible spin 
configurations $Z^r_i=Z^l_j=Z^{lr}_i$. A two-bond move cannot always be performed in this combined space, since the re-configured bonds may not be compatible with 
the current spin configuration. Starting with an allowed combined configuration $(V_l,V_r,Z^{lr}_i)$ (an example of which is shown in Fig.~\ref{fig1}), 
we carry out $N/2$ two-bond attempts in each of the bond configurations $V_l,V_r$, and thereafter construct all VB loops and flip each loop (i.e., flip all the 
spins in the loop) with probability $1/2$. The number of operations required for one such full cycle of updates scales as $N$. In the scaling of the computational 
effort we thus avoid the factor $\langle L_{\rm loop} \rangle = N \langle M_s^2\rangle $.
If there is antiferromagnetic order (system spanning-loops, $L_{\rm loop} \propto N$), or even in the absence of long-range order if the correlation 
length is large, the time savings of switching to the combined spin-bond basis is very large.

It is worth pointing out the reason why this method of avoiding to compute the overlap (\ref{overlap1}) works. It is because the combined Monte Carlo sampling 
of spins and bonds will, according to the standard detailed balance and ergodicity principles, automatically favor bond configurations for which there are 
more compatible spin configurations, on average exactly in proportion to the re-written overlap, Eq.~(\ref{overlap2}), which is just the number of spin 
configurations compatible with a given bond configuration. Thus, the overlap is taken into account probabilistically in the extended space, instead of being 
computed exactly in the pure VB space.\cite{evertz_Review} One may ask whether this could lead to a worse performance of the combined bond-spin sampling relative to pure VB 
sampling. The bond sampling in the mixed space is constrained, but, on the other hand, in the pure VB sampling a decreasing overlap after a bond 
reconfiguration (which should occur in half of the updates) reduces the acceptance rate. These two effects should, on average, balance each other, and 
thus the performance of the two methods should be similar when measured in terms of the number of bond updates performed. The sampling is, however, 
much faster in the combined space. 

Note also that measurements of observables can (and normally should) still be carried out in the pure VB basis (instead of measuring just the $z$-components 
of, e.g., correlation functions), i.e., at this stage it does not matter what the spin configuration is, and one just considers rotationally invariant estimators, 
such as Eq.~(\ref{ssestimator}), in the pure VB basis. This corresponds exactly to summing over all compatible spin configurations, i.e., one can consider the
VB (loop) estimators as {\it improved estimators}.\cite{evertz_Review}

\subsection{Bond-loop updates}

\begin{figure}
\center{\includegraphics[width=4.5cm, clip]{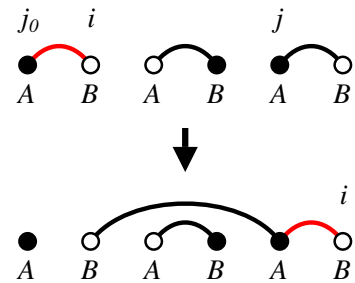}}
\vskip-2mm
\caption{(Color online) The first step in a bond-loop update. Open and solid circles indicate up and down spins compatible with the valence bonds. Given 
an initial site $j_0$, the site $i$ linked to it is identified. The bond $j_0,i$ is then replaced by $i,j$, with $j$ on the opposite sublattice 
from $i$ and chosen probabilistically as discussed in the text. If $S^z_i = S^z_j$, the new bond configuration is not allowed, and a new $j$ should 
be generated. The bond move results in one site with no bond and one with two bonds (unless $j=j_0$ and the loop update is completed). The site linked 
by the old bond to the site with two bonds becomes site $i$ for the following step. These procedures are repeated until the loop closes ($j=j_0$).} 
\label{fig3}
\vskip-3mm
\end{figure}

As an alternative to sampling the VB states using the two-bond reconfigurations, one can also implement a loop update similar to one developed for classical 
dimer models.\cite{adams,sandvikdimer} The idea is to first move one end of a randomly chosen bond, thereby creating two defects (an empty site and one site with
two bonds) that should not be present in a valence-bond state. Subsequent bond moves are then carried out to move one of these defects, until it annihilates 
the second one (the loop closes) and a new allowed bond configuration has been generated. The first step of such a loop update is illustrated in 
Fig.~\ref{fig3}. (Note that these loops are different from those in the rest of the paper).

In an algorithm (as well as in an actual computer implementation), it is convenient to represent the bonds by links between sites, stored in a one-dimensional 
array $v(i)$, $i=1,\ldots,N$. Thus, if sites $i$ and $j$ are connected by a bond, then $v(i)=j$ and $v(j)=i$. Starting at a randomly chosen site $j_0$ (in either 
$|V_l\rangle$ or $|V_r\rangle$; here we consider $|V_r\rangle$ with the links stored in an array $v_r$) the bond connected to it is considered for a reconfiguration. 
The bond will stay attached to its second site, $i=v_r(j_0)$, and the new other end, $j$, will be chosen according to probabilities proportional to the corresponding 
amplitudes $h({\bf r}_{ij})$. This is accomplished in practice by carrying out a bisection search for values bracketing a random number in $[0,1)$ in a pregenerated 
table containing cumulative probabilities, i.e., normalized partial sums of the amplitudes.\cite{longrange} If, for the chosen site $j$ [obtained by adding the 
chosen bond vector ${\bf r}$ to the current site $i$] the spin $S^z(j)=S^z(i)$, the move would not be consistent with the spin configuration, and a new $j$ should 
then be chosen [repeatedly if necessary, until $S^z(j)\not=S^z(i)$ is satisfied]. After moving the bond to an acceptable $j$, setting $v_r(i)=j$, this site will 
have two bonds on it (but in the array $v_r$ the original link has been destroyed, which does not matter since it will no longer be needed), and the original site 
$j_0$ will have no bond attached to it (unless the chosen site happens to be the starting site, $j=j_0$, in which case the loop update is immediately terminated 
and another loop building is started from a new randomly chosen site \cite{dimernote}). To ``heal'' the double-bond defect at $j$, the end of the old bond connected 
to it is moved. This is done in the same way as for the initial bond move, with $j_0$ in the description above replaced by $j$. The procedures are repeated until it 
happens that $j=j_0$, at which point the double-bond and no-bond defects annihilate each other and the loop closes. The loops can be large, and this kind of 
update should therefore be much more efficient than the local two-bond updates. The number of loops constructed in each updating sweep should be adjusted so that, 
on average, a number $\propto N$ of bonds are moved.

For an amplitude-product state with non-zero amplitudes $h({\bf r})$ for all ${\bf r}$, the bisection search in the bond-loop update introduces a factor 
$\propto \ln(N)$ in the time scaling of the algorithm, which may or may not (depending on the nature of the state) be outweighed by shortened autocorrelation times 
relative to the two-bond updates. Note also that if the amplitudes are zero or very small beyond some distance $r$ (e.g., in the case of exponentially decaying 
amplitudes), the table of probabilities should only include those amplitudes with practically non-vanishing amplitudes, which also removes the $\ln(N)$ factor 
from the scaling in such cases.

\section{Projector Monte Carlo}
\label{projector}

We now turn to the projector QMC method, using the Heisenberg model as a concrete example of loop updates for efficient ground-state projection of a 
variational state in the VB basis. We write the hamiltonian on a bipartite lattice as
\begin{equation}
H = -\sum_{\langle a,b\rangle}H_{ab},~~~ H_{ab} = -({\bf S}_{a} \cdot {\bf S}_{b}-\hbox{$\frac{1}{4}$}), 
\label{ham}
\end{equation}
where $\langle a,b\rangle$ denotes nearest-neighbor sites. The bond operators $H_{ab}$ are singlet projectors (equivalent to loop operators,\cite{evertz_Review} 
up to a factor of $2$ and bipartite rotation) which can have two effects when acting on a VB state:
\begin{eqnarray}
&&H_{ab}(a,b)=(a,b), \label{habdia} \\
&&H_{ad}(a,b)(c,d)=\half (a,d)(c,b). \label{haboff}
\end{eqnarray}
These simple rules, and the absence of minus signs (in the case of an unfrustrated system), 
enable a QMC scheme for projecting out the ground state $|\Psi_0\rangle$ from an arbitrary state 
(\ref{psivb}) in the VB basis.\cite{liang2,vbmethod1,vbmethod2} 

\subsection{Ground state projection} 

Irrespective of the basis and hamiltonian, the projector approach is based on the expression
\begin{equation}
(-H)^m |\Psi\rangle = c_0(-E_0)^m\left [|0\rangle + \sum_{n=1}^{\Lambda-1}\frac{c_n}{c_0}\left (\frac{E_n}{E_0}\right )^m |n\rangle \right ], 
\end{equation}
where $|n\rangle$, $n=0,\ldots,\Lambda-1$ are the energy eigenstates in a Hilbert space of $\Lambda$ states. If $E_0$ is the largest (in magnitude)
eigenvalue and the expansion coefficient $c_0\not =0$, then $|0\rangle \propto (-H)^m |\Psi\rangle$ for large $m$. In a QMC projector scheme, a high
power of $H$ and its action on the trial state $|\Psi\rangle$ are sampled stochastically.

To this end, for the Heisenberg model we write $(-H)^m$ as a sum of all products of $m$ bond operators and introduce the notation $P_\alpha$ for 
such operator strings;
\begin{equation}
(-H)^m = \sum_{\alpha} \prod_{i=1}^m H_{a^\alpha_i b^\alpha_i} = \sum_{\alpha} P_\alpha,
\label{hm}
\end{equation}
where $\alpha$ is a formal label for the different strings of singlet projectors. We write the state resulting when acting with a $P_\alpha$ on a given VB 
state $|V_r\rangle$ as
\begin{equation}
P_\alpha|V_r\rangle = (\half)^{o^r_\alpha}|V_r(\alpha)\rangle,
\label{propagate}
\end{equation}
where $|V_r(\alpha)\rangle$ is obtained in practice by successively applying the rules (\ref{habdia}) and (\ref{haboff}), which also gives the number
$o^r_\alpha$ of off-diagonal operations (\ref{haboff}). The expectation value of an arbitrary operator $O$ can be written as
\begin{equation}
\null\hskip-2mm
\langle O \rangle = \frac{\langle \Psi | (-H)^m O (-H)^m |\Psi\rangle}{\langle \Psi|(-H)^{2m} |\Psi\rangle} =
\frac{\sum_{lr\alpha\beta}W^{\alpha\beta}_{lr}O^{\alpha\beta}_{lr}}
{\sum_{lr\alpha\beta}W^{\alpha\beta}_{lr}},
\end{equation}
where the weight $W^{\alpha\beta}_{lr}$ and estimator $O^{\alpha\beta}_{lr}$ are
\begin{eqnarray}
W^{\alpha\beta}_{lr} &=& w_l w_r \langle V_l(\beta) |V_r(\alpha)\rangle 2^{-(o^\alpha_r+o^\beta_l)}, \label{vbweight} \\
O^{\alpha\beta}_{lr} &=& \langle V_l(\beta) | O |V_r(\alpha)\rangle/\langle V_l(\beta) |V_r(\alpha)\rangle. \label{vbestim}
\end{eqnarray}
The expectation value can be evaluated by importance-sampling, as discussed in Ref.~\onlinecite{vbmethod1}. However, up until now the sampling was rather 
inefficient. Each random replacement of operators (one or several; the number is adjusted to give a suitable acceptance rate) in $P_\alpha$  requires a full 
propagation of the current VB state, counting $o^r_\alpha$ in Eq.~(\ref{propagate}), and thereafter counting $N_{\rm loop}$ in the overlap (\ref{overlap1}). That is,
unlike what is normally the case in Monte Carlo simulations, one here cannot simply determine the weight ratio locally when a small change is made in the 
configuration, and, hence, the full weight has to be computed in each update. This results in a scaling $\sim {\rm max}(m^2,Nm)$ of the computational effort 
involved in a full updating sweep. Here one factor $m$ corresponds to the effort of propagating the full updated operator string. Constructing and
counting the loops in the transposition graph (\ref{overlap1}) requires $\propto N$ operations. If $m\gg N$ this effort is swamped by the bond propagation, 
and the total effort of carrying out a number $\propto m$ of updates (needed to significantly change the operator sequence) is $\propto m^2$, whereas for the 
situation $N>m$ (which is less interesting in practice, as we will see when discussing the convergence properties), the effort is formally $\propto Nm$. When 
using an amplitude-product as the trial state (instead of a fixed bond configuration), we should also perform $\propto N$ two-bond updates of the trial state, 
each of which demands the same computational effort as an operator replacement.

\begin{figure}
\includegraphics[width=6.75cm, clip]{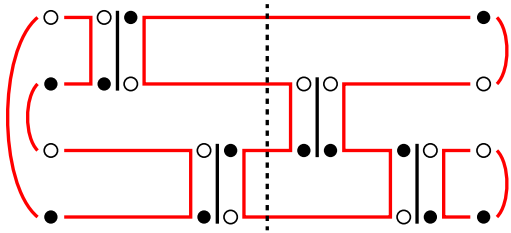}
\vskip-2mm
\caption{(Color online) A VB-spin-operator configuration contributing to $\langle \Psi |(-H)^{2m} |\Psi\rangle$ for a 4-site system with $m=2$. The arcs
to the left and right indicate VB states $\langle V_l|$, $| V_r \rangle$ and the two columns of filled and open circles represent $\up$ and 
$\dn$ spins of compatible spin states $\langle Z^l_j|$, $| Z^r_j \rangle$. The spins at the four operators (vertices) are also indicated. 
There are three loops, part of which consist of VBs. Expectation values are evaluated at the mid-point indicated by the dashed line.}
\label{fig4}
\vskip-3mm
\end{figure}

\subsection{Combined bond-spin space}

We will now show how using the combined spin-bond basis brings the scaling of one updating sweep down from ${\rm max}(m^2,Nm)$ to $\sim {\rm max}(N,m)$. We 
construct a loop update, which we here implement for the power $(-H)^{2m}$ in a way similar to the ``operator-loop'' update for ${\rm e}^{-\beta H}$, where 
$\beta=1/T$, in the SSE representation \cite{sse3} in $T>0$ simulations. The operator-loop approach is analogous to the original loop method for world lines, 
\cite{evertz_Review} with the main difference that the sampling scheme is formulated in terms of insertions and removals of operators, instead of deforming
world-line configurations. The formal relationships between the two approaches are discussed in Ref.~\onlinecite{evertz_Review}. The SSE formulation is 
somewhat easier to relate directly to the present case of projector QMC. We could also project with ${\rm e}^{-\beta H}$, Taylor-expanded as in the SSE method, 
but the fixed-power approach allows for some minor simplifications. The primary differences with respect to finite-$T$ SSE simulations is then the fixed value of 
$m$ and the VB ``end-cap'' boundaries in the ``propagation'' direction,  replacing the periodic boundary conditions appropriate at finite $T$. 

To make the analogy with the SSE loop method as close as possible, we split the singlet-projector operators $H_{ab}$ into their diagonal, $H_{ab}(1)$, and 
off-diagonal, $H_{ab}(2)$, parts,
\begin{eqnarray}
&& H_{ab}(1)=(\hbox{$\frac{1}{4}$} - S^z_aS^z_b). \label{hab1} \\
&& H_{ab}(2)=-\hbox{$\frac{1}{2}$}(S^+_aS^-_b + S^-_aS^+_b). \label{hab2}
\end{eqnarray}
We use a superscript $e$ on the operator string $P^e_\alpha$ in (\ref{hm}) to denote the $2^m$ different combinations of diagonal and off-diagonal operators 
for a given full-operator string $P_\alpha$. With spins $Z^r_i$, $Z^l_j$ compatible with $V_r$, $V_l$, we sample VBs, spins, and operators, according to the weight
\begin{equation}
W^{\alpha\beta,ef}_{lr,ij}=w_l w_r (\half)^{2m+N/2}\propto  w_l w_r,
\label{weight}
\end{equation}
under the condition $P_\alpha^e|Z^r_i\rangle = P_\beta^f|Z^l_j\rangle$. The constraints exactly compensate for the other factors in the original weight (\ref{vbweight}) 
and there is no explicit dependence in (\ref{weight}) on the operator-string ($\alpha,\beta,e,f$) and spin ($i,j$) indices. An example configuration is shown in 
Fig.~\ref{fig4}. On a bipartite lattice the weights are positive, since minus signs present in the states (\ref{vrz}) compensate those arising from an odd number 
of off-diagonal operators (\ref{hab2}) (or, equivalently, all signs could be eliminated by a sublattice rotation \cite{evertz_Review}). 

\subsection{Sampling method}

We now briefly describe the Monte Carlo sampling procedures. Starting with VB configurations $V_r,V_l$ (where normally one would take $V_r=V_l$ for simplicity) 
and compatible spin configurations $Z^r=Z^l$, an initial string cotaining only diagonal operators $H_{ab}(1)$ can be used (consistent with the constraint that 
each operator must act on two anti-parallel spins). Successive configurations maintaining the constraints are generated with three types of updates. 

In the first update---the ``diagonal update''---the combined string $P_{\alpha\beta}^{ef}=(P_\beta^f)^TP_\alpha^e$ (where the transpose $T$ of an operator sequence just 
corresponds to writing it in the reverse order, corresponding to acting with it on a bra state instead of a ket) containing $2m$ operators is traversed and each 
diagonal operator in it is updated (moved to a randomly selected bond), under the condition that it acts on anti-parallel spins. This step corresponds to changing the 
vertex break-up in the original world-line loop scheme. \cite{evertz_PRL,evertz_Review} As in the SSE method,\cite{sse3,sse1} the constraints are checked by keeping 
the single state $Z(p-1)$, which is needed for moving a diagonal operator at location $p$ in the string. This state is obtained by acting on the originally stored 
ket spin configuration $Z(0)=Z_r$ with the first $p$ operators in the sequence. It is changed (by flipping two spins) whenever an off-diagonal 
operator is encountered in the course of 
traversing the positions $p=1,\ldots,2m$. At the end of this procedure the stored bra state is obtained, $Z(2m)=Z_l$, for a valid configuration. 

In a second updating stage---the loop update---a linked list of operator-vertices is first constructed. A vertex consists  of the spin states  ``entering'' and 
``exiting'' an operator, as shown in Fig.~\ref{fig4}. They connect, forming loops. The only difference with respect to the operator-loops in the SSE method is that a 
loop can now be connected to the ket or bra VB state, and the valence bonds constitute parts of such loops (replacing the periodic boundary conditions used at $T>0$).
To keep nonzero (indeed, constant) matrix elements of the operators $H_{ab}$, all spins on a loop have to be flipped together, in the process changing also 
$H_{ab}(1) \leftrightarrow H_{ab}(2)$. Each loop is flipped with probability $1/2$. In practice, all loops are constructed, and the random decision of whether
or not to flip a loop is made before the loop is constructed. Vertices in a loop that is not to be flipped are just flagged as visited, so that the same
loop is not traversed more than once (i.e., a loop construction is always started from a vertex-leg that has not yet been visited). 

The reason for constructing all the clusters and flipping each with probability $1/2$, instead of generating single clusters starting from random seed
locations and flipping them with probability $1$ (as in the classical Wolff method \cite{wolff}), is that the {\it de facto} loop structure is only changed when 
performing the diagonal updates. One would therefore potentially generate the same cluster several times, which would lead to lower efficiency compared to uniquely
identifying all clusters and flipping each at most once. In principle one could modify the algorithm with combined diagonal and cluster updates, but this is 
more complicated and would probably not lead to improvements in efficiency in most cases.

A flipped loop including one or several VBs will cause spin flips in the stored spin configurations $Z^l$ or $Z^r$. In the loop updating
procedure we do not have to explicitly keep track of any other spins than those in $Z^l$ and $Z^r$. The four spins at 
the operators (the vertex legs) are irrelevant at the loop updating stage, because all the vertices automatically
involve only operations on anti-parallel spins, both before and after a loop flip. For each vertex encountered when constructing a loop, we therefore 
simply have to change the operator-type index, $1 \leftrightarrow 2$, in the list of operators (i.e., the same list $P_{\alpha\beta}^{ef}$ used in the diagonal 
update and to construct the linked vertex list).

The third type of update---the state update---is identical to the VB reconfigurations described in Sec.~\ref{variational} for the variational calculation. Normally 
one would use an amplitude-product state with coefficients (\ref{wamp}), which enter in the weight (\ref{weight}). Reconfigurations of the bonds can be carried out 
with either two-bond or bond-loops moves, as explained in Sec.~\ref{variational}. They only change the loop connections at the VB ``end caps''.

\subsection{Measuring observables}

When measuring operator expectation values one can go back to a pure VB (=loop) representation, using the estimator (\ref{vbestim}). This corresponds to summing 
over all loop orientations. Most quantities of interest can be expressed in terms of the loops in the transposition graph corresponding to 
$\langle V_l(\beta) |V_r(\alpha)\rangle$.\cite{liang1,sutherland,beach,evertz_Review} Note that these transposition-graph loops can also be obtained from the ``space-time''
loops constructed in the updates, by connecting the sites (in practice just assigning a label, the loop number $\lambda_i$) crossed by the same loop at the 
propagation midpoint (indicated by a dashed line in Fig.~\ref{fig4}). 
The ''space-time'' loops can also provide access to imaginary time correlation functions \cite{evertz_Review} in the ground state (see section \ref{sec:discussion}).
Since there are no differences in the measurement procedures for equal time observables with respect to the original 
VB projector algorithm, we refer to the literature for this aspect of the simulations.\cite{vbmethod1,vbmethod2,beach}

In some applications, instead of measuring a ground state expectation value $\langle 0 |A| 0\rangle$ one is interested in matrix elements of the form 
$\langle R |A| 0\rangle$, where $|R\rangle$ is a reference state, normally the N\'eel state in the $z$-component basis. This corresponds to sampling the wave
function itself (generating the basis states with probability proportional to the positive-definite wave-function coefficients). The energy (including excitation 
energies in different momentum sectors) can be computed like this,\cite{vbmethod1,vbmethod2} and also calculations of entanglement entropy can be formulated in this 
way.\cite{alet,chhajlany,hastings} A mixed matrix element can also easily be sampled in the spin-bond basis. In this case the loops terminating on the state 
$|R\rangle$ should never be flipped, because $|R\rangle$ is a single spin configurations (in the case of the N\'eel state---other reference states are also 
possible and would require other rules for the boundary loops).

\section{Results}
\label{results}

As a demonstration of the efficiency of the methods, we present results for the sublattice magnetization $M_s$ of the 2D Heisenberg model. This quantity 
has been calculated in numerous previous studies, but the currently best published estimate, $M_s=0.3070(3)$, obtained on the basis of $T \approx 0$ QMC results 
for $L$ up to $16$, is already more than ten years old.\cite{sse2} Recently, the density matrix renormalization group method was used to calculate $M_s$ 
on rectangular lattices with $N \approx 200$ sites, giving a result consistent with the above value and with a similar precision.\cite{white} Results have 
also been obtained using finite-$T$ data and scaling forms that in principle allow simultaneous $T \to 0,L\to \infty$ extrapolations. With $L$ up to $160$ 
and $1/T$ up to $12$, Ref.~\onlinecite{beard2} reported $M_s=0.30793(3)$. This is higher than (and well outside the error bars of) the $T=0$ results cited above. 
In order to resolve the discrepancy, it would be useful to have ground state results based on larger lattices. Here we consider $L$ up to $256$. 

Below we first discuss convergence aspects of the VB method, including the behavior with different trial states, and then present results and finite-size 
extrapolation of the sublattice magnetization.

\subsection{Variational calculations}

\begin{figure}
\includegraphics[width=7cm, clip]{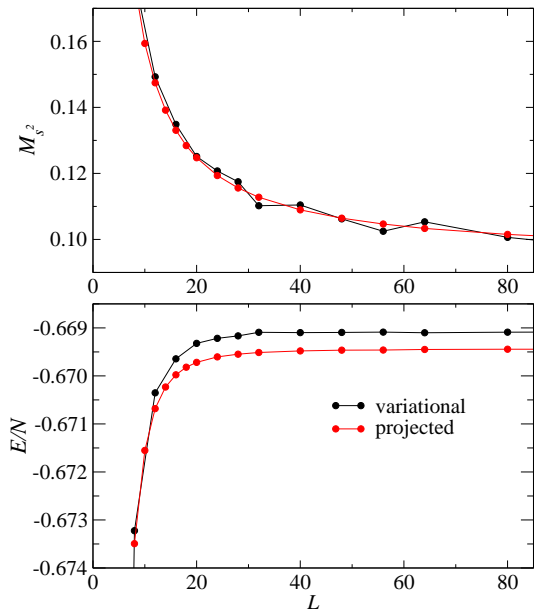}
\vskip-2mm
\caption{(Color online) The energy (lower panel) and the squared sublattice magnetization (upper panel) of the variational and 
ground-state projected states.}
\label{fig5}
\vskip-3mm
\end{figure}

We first discuss the amplitude-product states used as trial states for the ground-state projection. The quality of the variationally optimized 
states [i.e., all amplitude $h(x,y)$ were determined by variational Monte Carlo simulations, as explained in Sec.~\ref{variational}] is illustrated in 
Fig.~\ref{fig5} for system sizes $L$ up to $80$. Results for up to $L=32$ were previously presented in Ref.~\onlinecite{jievar}---here we improve
slightly on those results, thanks to the more efficient sampling procedures allowing for better statistics for the computed derivatives. The results 
are compared to converged results of the QMC projector method (which can be considered as exact to within small statistical errors that are not visible
in the graphs). The relative error of the variational energy is $<0.1\%$ for large systems. The sublattice magnetization falls on a a smooth curve 
in good agreement (better than 1\%) with the projected data for $L$ up to $\approx 24$. For larger systems the behavior becomes erratic, however, being higher 
or lower (outside the error bars) than the projected data in a seemingly random way. This can be explained as due to the energy becoming less sensitive to the 
long-range spin correlations for increasing $L$, i.e., there are states with significantly different sublattice magnetizations but energy expectation values 
that are the same to within the precision of the simulations. To obtain the correct best sublattice magnetization for large $L$ (corresponding to the minimum 
energy determined to extreme precision)  with the variational approach therefore requires unreasonably long simulations (which is true in general in 
variational calculations; not just with the amplitude-product states used here).

\subsection{Convergence of the ground-state projection}

\begin{figure}
\includegraphics[width=7cm, clip]{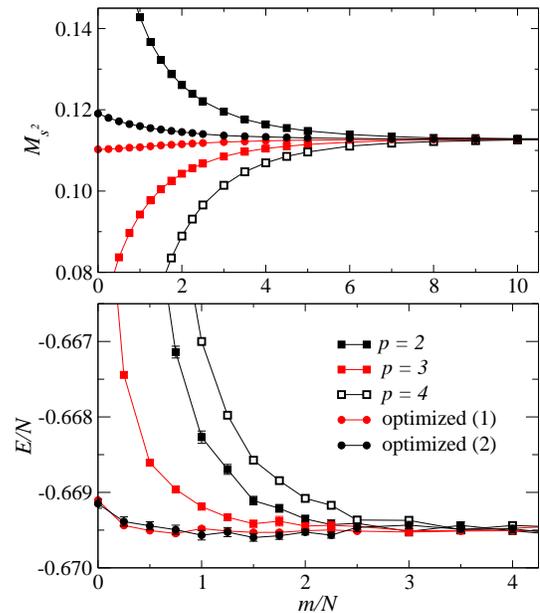}
\vskip-2mm
\caption{(Color online) Convergence of the energy (lower panel) and the squared sublattice magnetization (upper panel) for $L=32$ states
states projected using different trial states; amplitude-product states with amplitudes $h(r)=1/r^p$ ($p=2,3,4$) as well as with $h(x,y)$ 
determined by minimizing the energy (in two independent optimizations, giving slightly different amplitudes).}
\label{fig6}
\end{figure}

Turning now to results of the projector method, it is useful to test the convergence as a function of the projection power $m$ for different trial
wave functions. Clearly, the preferred option is to use the best variational state available, but optimizing an amplitude-product state also takes some 
time (depending on how close to the energy minimum one strives), and, as we have seen above, for large systems it may not even be possible to find the truly 
optimal amplitudes. Fig.~\ref{fig6} shows the energy and the sublattice magnetization for $L=32$ versus $m/N$, obtained using trial states with 
amplitudes $h(r)=1/r^p$, $p=2,3,4$, without any optimization, as well as with amplitudes obtained in two independent optimization runs. It is 
known \cite{jievar,beach2} that the optimal amplitudes decay as $1/r^3$ asymptotically, but the short-bond amplitudes show deviations from this form. 
Indeed, the best convergence is seen for $p=3$, but with optimized amplitudes the convergence is still much faster. Although the two optimized variational 
states have very similar energies, there are still clear differences in the convergence of the sublattice magnetization, related to the insensitivity of 
the variational energy to the long-distance spin correlations. 

\begin{figure}
\includegraphics[width=7.5cm,clip]{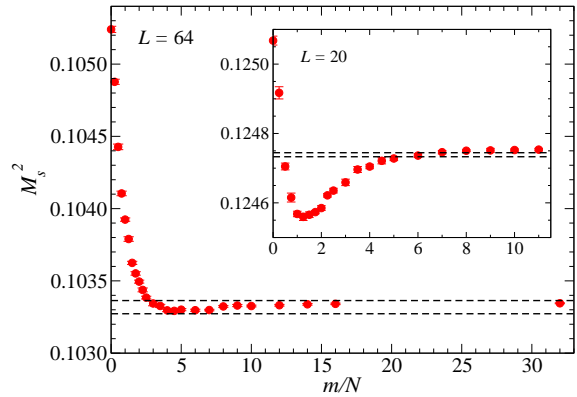}
\vskip-2mm
\caption{(Color online) Convergence of the squared sublattice magnetization for $L=64$ ($L=20$ in the inset), using an optimized trial state.
The dashed lines show the result $\pm$ error bar of SSE calculations (using loop updates) at very low temperatures ($\beta=8192$ in the case of $L=64$).}
\label{fig7}
\vskip-3mm
\end{figure}

In some cases the convergence of the sublattice magnetization is non-monotonic (while the energy always has to converge monotonically), as illustrated in 
Fig.~\ref{fig7}. The behavior depends on details of the variationally optimized amplitudes; likely non-monotonicity can be traced to incomplete optimization. 

\subsection{Extrapolation of the sublattice magnetization}

\begin{table}
\begin{center}
  \begin{tabular}{|c|l|l|}
    \hline
    ~~~$L$~~~    &  ~~~~~~~~$M^2_s$~~~~~~~~        & ~~$C(L/2,L/2)$~~ \\ \hline
8    &   ~~0.177843(1)    &    ~~0.137595(2)   \\
10   &   ~~0.159372(2)    &    ~~0.128552(2)   \\
12   &   ~~0.147448(2)    &    ~~0.122586(2)   \\
14   &   ~~0.139153(2)    &    ~~0.118380(2)   \\
16   &   ~~0.133067(2)    &    ~~0.115263(2)   \\
18   &   ~~0.128412(2)    &    ~~0.112857(2)   \\
20   &   ~~0.124748(2)    &    ~~0.110954(2)   \\
24   &   ~~0.119350(2)    &    ~~0.108125(2)   \\
28   &   ~~0.115573(2)    &    ~~0.106126(2)   \\
32   &   ~~0.112782(2)    &    ~~0.104636(2)   \\
40   &   ~~0.108943(3)    &    ~~0.102571(3)   \\
48   &   ~~0.106431(3)    &    ~~0.101208(3)   \\
56   &   ~~0.104661(3)    &    ~~0.100239(3)   \\
64   &   ~~0.103345(3)    &    ~~0.099514(4)   \\
80   &   ~~0.101523(4)    &    ~~0.098501(4)   \\
96   &   ~~0.100325(5)    &    ~~0.097831(5)   \\
128  &   ~~0.098843(16)   &    ~~0.096990(17)  \\
192  &   ~~0.097371(11)   &    ~~0.096161(11)  \\
256  &   ~~0.096669(17)   &    ~~0.095765(16)  \\
    \hline
\end{tabular}
\end{center}
\vskip-3mm
\caption{Projector QMC results for the squared sublattice magnetization and the spin correlation function at
maximal separation for several $L\times L$ lattices. The numbers within brackets indicate the statistical error (one
standard deviation of the average) in the last digit.}
\label{restab} 
\vskip-3mm
\end{table}

\begin{figure}
\includegraphics[width=8cm, clip]{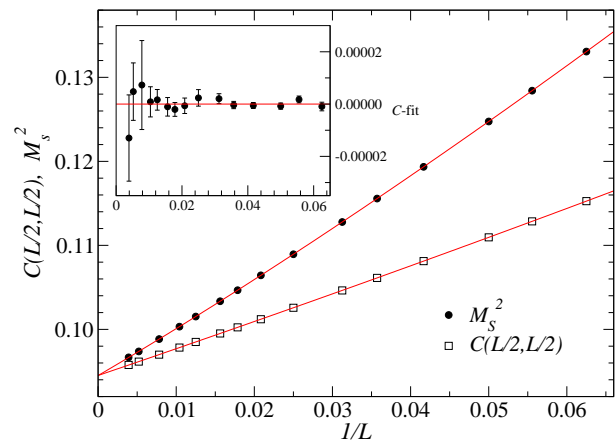}
\vskip-2mm
\caption{(Color online) Finite-size scaling of the sublattice magnetization. The curves are polynomials fitted to $16 \le L\le 256$ data (cubic
for $C$ and 4th-order for $M_s^2$). The inset shows the deviation of the simulation results for $C(L/2,L/2)$ from the corresponding fit.}
\label{fig8}
\vskip-3mm
\end{figure}

We now discuss large-scale calculations for the 2D Heisenberg model. We have calculated $M_s^2$ as well as the spin correlation function $C(L/2,L/2)$, 
which equals $M_s^2$ when $L \to \infty$, for lattices with $L$ up to $256$, making sure that the results are well converged to the ground state in all 
cases. The raw data are listed in Table.~\ref{restab} The results are graphed versus $1/L$ in Fig.~\ref{fig8}, along with polynomial fits \cite{reger} 
used to extrapolate to $L=\infty$. The extrapolated  $M_s^2$ and $C(L/2,L/2)$ agree statistically and are stable with respect to the range of $L$ included 
and the order of the polynomials. The statistics is slightly better for $C$ and the polynomial needed to fit it is one order smaller than for $M_s^2$. 
Based on $C$, we estimate $M_s=0.30743(1)$, somewhat above the previous $T=0$ results.\cite{sse2,white} The error bar is more than an order of magnitude 
smaller. The higher value from finite-$T$ simulations\cite{beard2} can be ruled out (differing by more than 15 of its error bars from our result). This 
illustrates difficulties with unknown corrections to the $(T,L)$ scaling forms. Extrapolating $T=0$ properties directly as a function of a single parameter 
($1/L$) can in general be expected to be more reliable. Indeed, since the appearance of the (unpublished) original short version of the present article,
\cite{arxiv} the results of Ref.~\onlinecite{beard2} have been recalculated and revised\cite{gerber} and now agree with our results.

\section{Summary and Discussion}
\label{summary}

We have shown how re-introducing the spins into the VB basis allows for much faster sampling of amplitude-product states and, in particular, very efficient 
ground-state projector simulations. For variational calculations it saves a factor up to system size $N$, and for projector QMC a factor up to the projection 
power $m$, bringing the total computational effort from $O(m^2)$ to $O(m)$ (where normally $m\gg N$, i.e., the improvements can be orders of magnitude). One 
striking and appealing aspect of the projector algorithm is its close similarity with state-of-the-art $T>0$ QMC loop methods, in particular the SSE variant 
\cite{sse3} but also the world-line approach (for which loop updates similar to those 
used here were  originally developed\cite{evertz_PRL,evertz_Review}). Essentially, the $T=0$ projector approach corresponds to ``cutting open'' the periodic imaginary-time 
boundary condition used at $T>0$ and terminating the resulting free loop ends with valence bonds (which act as continuations of the loops). It is
therefore very easy to rewrite, e.g., an SSE program for $T=0$ projection. A favorable aspect of the $T=0$ projector approach is that VB amplitude-product 
states often are very good trial states (as noted already a long time ago \cite{liang1,liang2}), which helps substantially to achieve a fast convergence 
as a function of the power $m$ of the projection operator $H^m$.

\subsection{Discussion} \label{sec:discussion}

\begin{figure}
\includegraphics[width=7cm, clip]{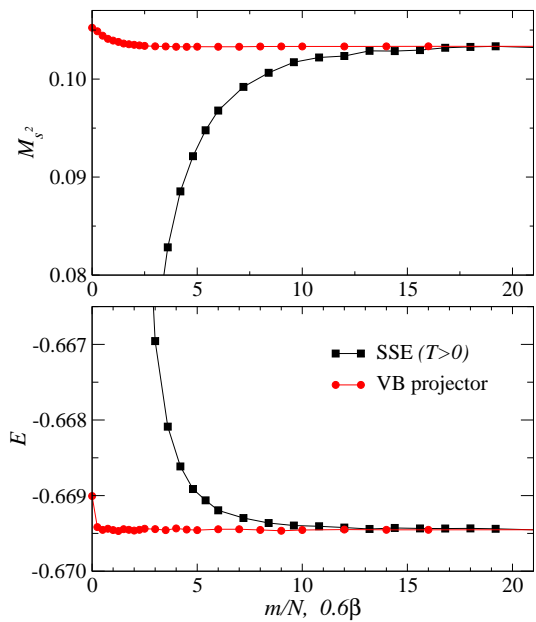}
\vskip-2mm
\caption{(Color online) Convergence of  VB projector results for the energy (bottom panel) and the sublattice magnetization (top panel) for
an $L=64$ system versus the power $m$ of the Hamiltonian used in the projection. The results at $m/N$ are compared with $T>0$ results graphed 
versus the normalized inverse temperature $0.6\beta$ (corresponding to a similar computational effort as the projection scheme at $m/N$).}
\label{fig9}
\vskip-3mm
\end{figure}

In view of the similarity with $T>0$ methods, it is worth asking how much faster a converged $T=0$ simulation is than a $T>0$ simulation carried out at $T$ 
sufficiently low for obtaining ground state properties. To answer this question, we first need to compare directly the size of the configuration space of the two 
formulations.  Consider the SSE method, which is based on sampling the Taylor expansion of ${\rm e}^{-\beta H}$.\cite{sse1}  The average of the expansion power $n$ 
is given by $\langle n\rangle = \beta |E|$, where $E$ is the total energy ($\propto N$), which includes for each bond (in the case of the $S=1/2$ Heisenberg model) 
an added constant $-1/4$ [exactly as in the singlet projector operator in Eq.~(\ref{ham})]. Let us consider the $2D$ Heisenberg model, where for large 
systems the energy per site is $\approx -0.67$, which, including the constant $-1/4$, corresponds to $|E|/N\approx 1.2$  Thus, the average expansion 
power $\langle n\rangle \approx 1.2N\beta$. In the projector approach, the number of operators in the sequence is $2m$, and thus the computational efforts 
of the two methods should be comparable if $m/N \approx 0.6\beta$ (since there is only a very small overhead in sampling in the projector scheme, related to 
the boundary VB states). In Fig.~\ref{fig9} we show results of the two methods versus these normalized 
parameters for $L=64$. A much faster convergence of the projector approach can be seen, especially for the energy. It is clear that here the fact that the
trial state is well optimized plays a big role---with a very poorly optimized trial state the $T>0$ approach may even converge faster. 

It should be noted that the loop estimators used in QMC calculations at $T>0$ are exactly equivalent to VB estimators,\cite{evertz_Review} such as 
Eq.~(\ref{ssestimator}), if $T$ is low enough for a simulation to sample only the singlet subspace, whereas at higher $T$ some loops---those that change the total 
magnetization when flipped---correspond to higher spin contributions.

 An important difference between the $T=0$ and $T>0$ approaches should be pointed out in this 
context: To improve the statistics in $T>0$ simulations one can take advantage of the periodic ``time'' boundaries. Averages of observables can be computed over 
all the propagated states. Whether or not this is a significant advantage in practice depends on the observable (its auto-correlation function as a function of 
imaginary time, which determines the effective number of independent measurements generated). In some cases, the averaging can be carried out almost without 
overhead. At first sight it might appear that no averaging of this kind can be done with the projector method, since the matrix elements of observables are defined 
at the mid-point of the configuration (as indicated in Fig.~\ref{fig4}). One can of course shift the measurement point, and average over several such points, although
then either the ket or the bra might not be equally well converged to the ground state. In practice, this kind of averaging may, however, still pay off, although we 
have not investigated it quantitatively. Note also that quantities given in terms of Kubo integrals (susceptibilities) explicitly require access to imaginary-time 
correlation functions, which can in principle be measured in the projector scheme by using the ''space-time'' loops constructed in the updates.\cite{evertz_Review} 
It would then be better to formulate using the Taylor expansion of
${\rm e}^{-\beta H}$ instead of a fixed power $(-H)^m$, and one would require longer projections than equal-time correlations, long enough for the 
imaginary-time correlator to have decayed below some small number beyond which further contributions can be neglected. These issues are common to projector
methods and have been investigated and discussed in detail in the context of fermionic auxiliary-field QMC calculations.\cite{furukawa,feldbacher}

\subsection{Related calculations and outlook}

The possibility to use the valence-bond basis in QMC simulations was first suggested more than 20 years ago.\cite{liang1} It is also implicitly the basis for 
the loop algorithm.\cite{evertz_Review} Other aspects of simulations explicitly formulated in the VB basis became more widely used 
only recently, after its generic utility, including simplified access to many physical quantities, was pointed out.\cite{vbmethod1,vbmethod2,beach}
We briefly list some works where the unique aspects of the VB basis where exploited in such simulations, and where the improvements presented here
can be expected to lead to further progress. 

Alet {\it et al.} \cite{alet} and, independently, Chhajlany {\it et al.}\cite{chhajlany} defined a VB entanglement entropy, which can be evaluated 
using ground-state projection (unlike other entropy definitions, which normally cannot be directly evaluated). A slightly different variant
of this entanglement measure was proposed by Lin and Sandvik, who also introduced another measure, the loop entanglement entropy, based on the
loops in the transposition graph.\cite{yucheng}  Hastings {\it et al.} recently showed that one of the standard definitions of entanglement entropy, the Renyi 
$S_2$ entropy, in fact also can be evaluated with VB simulations.\cite{hastings} These developments will allow, e.g., tests of the ``area law'' of 
entanglement entropy and deviations from it in a wide range of quantum spin systems. 

Beach  {\it et al.} have extended the projector scheme for standard
SU$(2)$ spins to an often used  representation of SU($N$) invariant models, including also the possibility of continuously varying $N$.\cite{beach3} 
This extension is of interest as many analytical theories are formulated as large-$N$ expansions, and it is useful to make direct contact between this 
approach and unbiased numerical calculations. Using a similar approach, Tran {\it et al.} have carried out VB simulations of a chain of non-abelian 
SU($2$)$_k$ particles (where $k$ is related to the central charge),\cite{tran} generalizing the standard $S=1/2$ Heisenberg chain. One can also 
consider effective hamiltonians explicitly constructed in the VB basis.\cite{lou2} 

One of the first applications of the pure VB projector method was to the J-Q model; a potential realization of ``deconfined'' quantum criticality.\cite{deconf} 
Since the appearance of the unpublished original short version of the present article,\cite{arxiv} the improved VB loop method has already been used to study 
SU($N$) versions of the J-Q model with $N=3$ and $4$.\cite{lou3} Here it can be noted that the variational amplitude-product states can also be extended to allow for 
valence-bond-solid order (by introducing factors favoring or disfavoring various local short-bond configurations), which is useful for studying J-Q models 
in the VBS phase.\cite{louthesis} 

The spin texture around a non-magnetic impurity in the J-Q model has been studied with the spin-bond sampling algorithm with an extra 
unpaired spin.\cite{damle} Triplet excitations of random antiferromagnetic clusters were investigated with VB simulations including a triplet VB in the
background of singlets by Wang and Sandvik,\cite{wang2} a measurement which also can be carried out with two unpaired parallel spins. Measurements of triplet 
excitations for fixed momentum were discussed in Ref.~\onlinecite{vbmethod2}.

Finally, we note that the loop-based projector scheme that we have presented here for isotropic spin systems can in principle be extended to anisotropic 
models as well\cite{evertz_Review,directed} (and also to more complex bosonic systems, as well as fermions in one dimensions). The difference will be that the variational trial states 
appropriate in most other cases would not have natural expressions in the singlet valence bond basis. As an example, consider the case 
of a variational wave function written just in terms of the $z$-components of the spins (e.g., a Jastrow state). There is then some weight (wave function 
coefficient) $W(\{S_i^z\})$ for the trial state, and normally this weight will change when a loop terminating at the trial state is flipped. The loops in 
this case would be of the ``directed'' type, \cite{directed} and one could build in the weight-change coming from the trial state into the directed loop 
equations (which dictate the probabilities for the loop-building along different paths through the vertices). This approach may be more efficient than the 
Green's function or diffusion Monte Carlo methods \cite{trivedi} normally used for ground-state projection of a variational trial state.
\null\vskip-9mm\null

\acknowledgments

We thank Kevin Beach for many useful discussions. AWS was supported by the NSF under Grants No.~DMR-0513930 and ~DMR-0803510. Part 
of this work was carried out by AWS while on sabbatical at the Institute for Solid State Physics, University of Tokyo, Japan, and at the Department 
of Physics, National Taiwan University, Taipei, Taiwan. He would like to thank these institutions for their hospitality and financial support.

\null


\begin{thebibliography}{00}
\bibitem{dimernote}
In Ref.~\onlinecite{sandvikdimer} a ``directed loop'' scheme was used for classical dimers with restriction to short bonds (of two different lengths). In that 
case there are no bond updates in which a bond does not move (i.e., in which the generated bond is the same as the current one). However, this approach is not practical 
when there is no restriction on the bond lengths (as is required in the over-complete VB basis). The probability of not moving a bond is low and in practice 
the scheme of bond selection from a single list with all bonds is still very efficient.
\end{thebibliography}
\end{document}